\documentclass[10pt,conference]{IEEEtran}
\IEEEoverridecommandlockouts
\usepackage{cite}
\usepackage{amsmath,amssymb,amsfonts}
\usepackage{algorithmic}
\usepackage{graphicx}
\usepackage{multirow}
\usepackage{float}
\usepackage{subcaption}
\usepackage{caption}
\usepackage{array}
\usepackage[skip=0.5ex]{subcaption}
\usepackage{xcolor}
\usepackage{braket}
\usepackage{hyperref}
\usepackage{booktabs}
\usepackage{fancyvrb}
\usepackage{dblfloatfix}

\def\BibTeX{{\rm B\kern-.05em{\sc i\kern-.025em b}\kern-.08em
    T\kern-.1667em\lower.7ex\hbox{E}\kern-.125emX}}

\newcolumntype{a}{>{\columncolor{gray!10!white}}c}
\newcolumntype{x}{>{\columncolor{green!10!white}}c}
\newcolumntype{y}{>{\columncolor{blue!10!white}}c}
\newcolumntype{z}{>{\columncolor{yellow!10!white}}c}
\newcolumntype{v}{>{\columncolor{red!10!white}}c}

\definecolor{OliveGreen}{rgb}{0,0.6,0}
\definecolor{ForestGreen}{RGB}{34,139,34}
\definecolor{myblue}{RGB}{37,165,203}
\definecolor{FAUblue}{rgb}{0.000, 0.2196, 0.3961}
\definecolor{myred}{RGB}{175,32,67}

\colorlet{backgroundcol}{cyan!10!white}

\newcommand{\jennifer}[1]{\textcolor{black}{#1}}

\begin{document}

\title{Quantum Computer Simulations at Warp Speed: Assessing the Impact of GPU Acceleration\\\thanks{Preprint submitted for publication.}
{\large A Case Study with IBM Qiskit Aer, Nvidia Thrust \& cuQuantum}
}
  
\author{\IEEEauthorblockN{Jennifer Faj, Ivy Peng, Jacob Wahlgren, Stefano Markidis}
\IEEEauthorblockA{\{faj,ivybopeng,jacobwah,markidis\}@kth.se\\\textit{Department of Computer Science}\\\textit{KTH Royal Institute of Technology, Sweden}}
}

\maketitle

\begin{abstract}
Quantum computer simulators are crucial for the development of quantum computing. In this work, we investigate the suitability and performance impact of GPU and multi-GPU systems on a widely used simulation tool -- the state vector simulator \texttt{Qiskit Aer}. In particular, we evaluate the performance of both Qiskit's default Nvidia \texttt{Thrust} backend and the recent Nvidia \texttt{cuQuantum} backend on Nvidia A100 GPUs. We provide a benchmark suite of representative quantum applications for characterization. For simulations with a large number of qubits, the two GPU backends can provide up to $14\times$ speedup over the CPU backend, with Nvidia cuQuantum providing further $1.5-3\times$ speedup over the default Thrust backend. Our evaluation on a single GPU identifies the most important functions in Nvidia Thrust and cuQuantum for different quantum applications and their compute and memory bottlenecks. We also evaluate the gate fusion and cache-blocking optimizations on different quantum applications. Finally, we evaluate large-number qubit quantum applications on multi-GPU and identify data movement between host and GPU as the limiting factor for the performance. 
\end{abstract}

\begin{IEEEkeywords}
GPU, State Vector Quantum Computer Simulator, \texttt{Qiskit Aer}, Performance Characterization
\end{IEEEkeywords}

\section{Introduction}
Accelerators and GPUs impacted critically both HPC applications and machine learning workloads. Outstanding advancement examples are the acceleration of molecular dynamics software~\cite{andersson2022breaking}, molecular docking~\cite{schieffer2023tcu}, computational fluid dynamics~\cite{karp2022large}, plasma codes~\cite{chien2020sputnipic}, and weather forecast~\cite{fuhrer2018near}. GPUs revived deep-learning applications after the long AI winter~\cite{krizhevsky2017imagenet}. In the current days, there is no efficient and large-scale deep neural network training without accelerators and specialized hardware. This study focuses on understanding whether GPU acceleration is a key technology for an emerging HPC application that is the usage of classical computing for simulating current and upcoming quantum computers. The question we want to answer in this work is: \emph{are GPUs suitable and a key enabling technology for the acceleration of quantum computer simulations?} %

Quantum computer simulators are a key tool for the development of quantum computing. In fact, the design and implementation of large-scale, reliable quantum computing infrastructure (hardware, software and algorithmic) require the deployment of quantum computer simulators. These tools can drive quantum computer design choices, prototyping quantum algorithms in ideal and noisy controlled environments, and verifying the correctness of the real quantum computer. In fact, most of the current quantum algorithmic development depends on prototyping the algorithms and running on quantum computer simulators to assess the correctness of the results in controlled environments and the impact of noise and error on the algorithms.

From the general stand-point, there exist two main categories for quantum computer simulators. The first category models quantum computer closer to the hardware and describes how to control pulses (typically microwave pulses in superconducting and trapped ion systems) to implement quantum circuit operations. These simulators go under the name of pulse-level simulators. Examples of these are the IBM OpenPulse~\cite{gokhale2020optimized} and PASQAL Pulser~\cite{silverio2022pulser}. The second approach is at higher-level and much more common in use. It uses abstractions, such as quantum gates and circuits. This class includes several simulation techniques, among which state vector (also called \textit{Schrödinger}), tensor network contraction~\cite{markov2008simulating}, and Feynman path~\cite{markov2018quantum} techniques are the most important ones. By far, currently the most established quantum simulator is the state vector quantum computer simulator because of its simplicity and easiness of implementation. Differently from other algorithms, state vector is not suitable for noisy simulations (density matrix is a convenient approach instead) and it is limited to a simulation with a relatively small number of qubits (the basic unit of information in quantum computing), e.g. $\le$ 48 qubits on the current largest HPC systems~\cite{wu2019full}. On the other hand, tensor networks can simulate hundreds to thousands qubits for low entangled networks but they do not provide the full state vector. The main limitation of state vector simulators is that memory usage scales exponentially with the number of qubits. For instance, a full state vector simulation requires the full memory of the Summit supercomputer, 2.8PB~\cite{wu2019full}. Because of these limitations, it is not clear whether the state vector quantum computer simulator might benefit from the usage of GPUs.

\begin{figure*}
    \centering
        \includegraphics[width=0.65\textwidth]{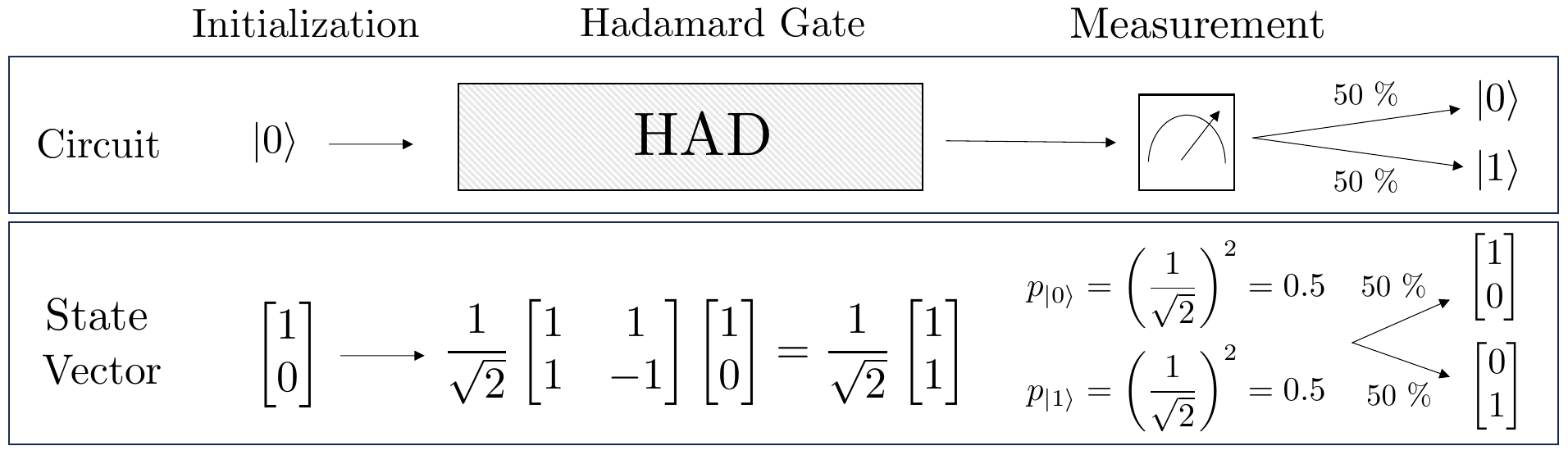}
        \caption{A state vector simulator of a one qubit circuit with an Hadamard gate and a measurement. The Hadamard gate corresponds to a matrix-vector multiply. A measurement corresponds to a random sampling of a state according to its probability.} 
        \label{fig:simulator}
\end{figure*} 

This work has the goal of evaluating the suitability of GPUs for state vector computer simulators and identifying opportunities for their deployment on accelerated systems. To answer our research question, we use the IBM Qiskit Aer state vector simulator \cite{QiskitCommunity2017}, a state-of-the art quantum computer simulator, providing two backends for Nvidia GPUs: one with Nvidia Thrust and one with  Nvidia cuQuantum.

The contributions of this paper are the following:
\begin{itemize}
\item We develop a benchmark suite of six representative quantum applications that support scaling the number of qubits in a state vector quantum computer simulator.
\item We characterize and compare the performance of the state-of-the-art Qiskit Aer quantum computer simulator with three backends including CPU, GPU with Nvidia Thrust, and Nvidia cuQuantum on Nvidia A100 GPUs.
\item We analyze the impact of two critical optimizations in transpilation -- gate fusion and cache blocking, on different quantum applications.
\item Our evaluation on a single GPU identifies the most important functions in Nvidia Thrust and cuQuantum for different quantum applications and we built the roofline to identify compute and memory bottlenecks. 
\item Our evaluation on a multi-GPU setup identifies data movement between host and GPU as the top one factor limiting the performance of large-number qubit quantum applications. 
\end{itemize}

\section{Quantum Computer State Vector Simulator}\label{sec:background}
\begin{figure*}
    \centering
        \includegraphics[width=\textwidth]{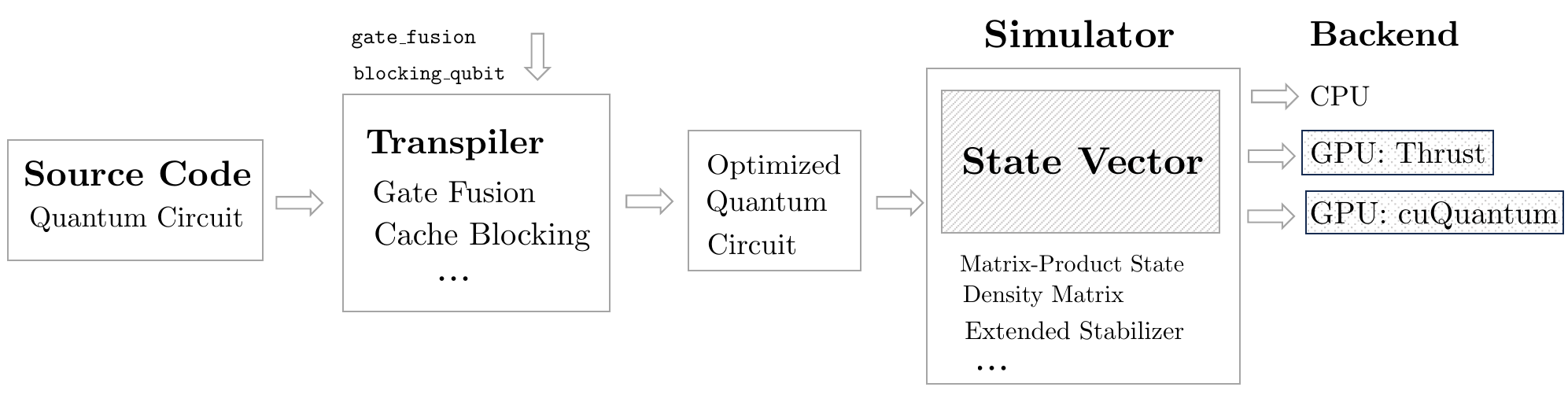}
        \caption{Qiskit Aer workflow when running a state vector simulations.} 
        \label{fig:workflow}
\end{figure*}
In this work, we focus on simulators using high level abstractions, such as quantum gates and circuits, and the so-called state vector approach that stores all the information about quantum states. \\

\noindent\textbf{Quantum Systems and Transformations with Complex Arrays and Matrices.} In quantum computing, the simplest quantum system is the quantum bit or \emph{qubit}. This is a two-state quantum system that can be expressed with a linear superposition of its two orthonormal basis states (the $\ket{0}$ and $\ket{1}$ states) as $\psi = c_0 \ket{0} + c_1 \ket{1}$, where  $c_0$ and $c_1$ are two complex numbers such that $\|c_0\|^2 + \|c_1\|^2 = 1$. In a state vector simulator, the qubit is represented by a two-element complex-value array as $[c_0 \: c_1]^T$. An important abstraction is the quantum gate that allows us to manipulate the state of a quantum bit. Each quantum gate must be reversible and conserve the total probability of the states to one, to obey to the quantum mechanics law. Examples of single qubit gates are the NOT (or X), rotation gates, and the most famous Hadamard gate to prepare an initial state into equal superposition of states. In state vector quantum computer simulators, each one-qubit gate is expressed as a $2\times2$ unitary matrix  (to enforce the reversibility and conservation of probability). For instance the Hadamard matrix is $\frac{1}{\sqrt{2}}\begin{bmatrix} 1 & 1 \\ 1 & -1 \end{bmatrix}$. For a comprehensive description of quantum gates and associated matrices, we refer to quantum computing textbooks~\cite{nielsen2002quantum}. A gate transformation on a qubit is expressed as a matrix-vector multiplication, where the matrix is a gate matrix and the vector is the input state vector. If two or more gates are in sequence, it is possible to calculate the final effect of the gate sequence, by simply multiplying the gate matrices. Because of this approach, the state vector calculations are dominated by small-size matrix-vector multiplications. In practice, quantum simulators do not store matrix information but they are implemented in place (matrix-free). As noted in the seminal paper on HPC state vector simulator~\cite{haner20175}, the basic computational building block of state vector simulator has relatively low arithmetic intensity (a matrix multiplication requires 14 FLOPs and 16B need to be moved when using single precision). 

When we measure the qubit, its quantum state $\psi$ collapses to either basis state of $\ket{0}$ or $\ket{1}$ with probabilities $p_{0} = \|c_0\|^2$ and $p_{1} = \|c_1\|^2 = 1 - \|c_0\|^2$. In a state vector simulator, the measurement outcome in a basis state is simply simulated with a random sampling using the probability $p_1$ and $p_2$ to decide the measurement outcome. Multiple measurements or shots can be modeled with several sampling using the same full state vector calculated by the simulator. Fig.~\ref{fig:simulator} shows an example of a simple state vector quantum computer simulation of a one-qubit circuit with an Hadamard gate. The qubit is initialized in a classical basis state ($\ket{0}$ in the Fig.~\ref{fig:simulator}), corresponding to a state vector $[1 \: 0]^T$. To apply an Hadamard gate, we multiply the Hadamard matrix to the input qubit. As result of this transformation, we obtain the state vector $\frac{1}{\sqrt{2}}[1 \: 1]^T$: the qubit is an equal superposition of the basis states $\ket{0}$ and $\ket{1}$.

In quantum mechanics, quantum systems are combined by using the tensor product. For instance, we can combine two qubit systems $q_1 = [c_0 \: c_1]^T$ and $q_2 = [c_2 \: c_3]^T$ and the result will be a multi-qubit system $q_{q_1 \otimes q_2} = q_1 \otimes q_2 = [c_0 c_2 \: c_0 c_3 \: c_1 c_2 \: c_1 c_3 ]^T$. From this simple example of combining two qubits, it is clear that combining $N$ qubit systems in one multiqubit system will require $2^N$ complex elements in the state vector simulator, leading to an exponential increase of the memory requirements with $N$. For instance, a simulation with 31 qubits requires $2^{31} \times 2 \times 4 / 10^9 \textrm{GB} = 17.2\textrm{GB}$ for the state vector array in single precision.

With multi-qubit systems, we can have one-qubit gate acting in parallel on different qubits: in this case the gate matrix acting on the full state vector can be calculated as the tensor product of the gate matrices. For instance, in the case of a two qubit system with two Hadamard gates, thanks to the tensor product, we can fuse two Hadamard matrices in one $4 \times 4$ matrix acting on the full state vector. In multi-qubit systems, we also have true multi-qubit gates, e.g. transformations that take as input the values of two or more qubits. The archetype of multi-qubit operations are the controlled operations, akin to classical conditional operations: a one-qubit gate/transformation is performed on a target qubit on the condition that a control qubit is in the state $\ket{1}$. An example of multi-qubit gate is the two-qubit C-NOT and three-qubit version of it, called CC-NOT or Toffoli gate. Controlled gates introduce entanglement. When it comes to state vector, multi-qubit operations are still one-qubit operations that act only on selected qubits as determined by the control qubits (this is because of the conditional nature of controlled operations).\\

\noindent\textbf{Qiskit Aer State Vector Simulator.} Qiskit is an open-source software stack, initially released by IBM in 2017, for developing codes for quantum computers at the level of circuits, pulses, and algorithms. The Qiskit Aer component provides a range of high-performance quantum computing simulators, including the state vector simulator we use in this work, and a GPU port by Doi et al.~\cite{doi2019,doi2020cache} using the Thrust library  and recently also a cuQuantum~\cite{the_cuquantum_development_team_2023_7806810} port. The open-source IBM Qiskit Aer programming framework~\cite{QiskitCommunity2017} provides an ideal environment to evaluate state vector performance on GPU and compare it with the CPU version. Fig.~\ref{fig:workflow} shows a high-level description of the workflow when running the IBM Qiskit Aer simulator. 

When running a quantum computer simulator, the first step is to provide a description of the quantum circuit to be executed. In quantum computing, algorithms and codes are expressed as quantum circuits with multi-qubit and a combination of quantum gates and measurements. In Qiskit Aer, this can be formulated in Python code using quantum gates, libraries of quantum gates (for instance the Quantum Volume and QFT circuits) or by loading circuit information from an input file in the QASM format~\cite{cross2017open}, a standard description of circuit in a style reminiscent of classical Assembly.

The second step is the Qiskit Aer transpilation step. This is a critical part of any quantum computing software stack and it is mainly responsible for circuit optimization and mapping to underlying quantum computer topology via several compiler passes. When using the state vector quantum computer simulator, the transpiler is responsible for performance optimization for increasing the computation intensity and minimizing the data movement. For our research, the two critical performance optimization transpiler passes are:

\begin{itemize}
\item \textit{Gate Fusion.} A key performance optimization technique in state vector simulator is to fuse two or more individual gate matrices into a fused matrix. For instance, we can fuse gates acting on different qubits by taking the tensor product of the individual gate matrices acting on different qubits. This will increase the computational intensity of the matrix multiply. It is also possible to fuse two or more gates acting on a single qubit by taking the matrix multiplication (and respecting the order since matrix multiply is not commutative). In Qiskit Aer, the parameter \texttt{fusion\_threshold} determines the threshold that the number of qubits must be greater than or equal to to enable the fusion optimization. The default threshold is 14 qubits.

\item \textit{Cache Blocking.} On multi-GPU systems and in general on distributed memory systems, we need to divide the state vector on different \textit{chunks} (to use Qiskit Aer terminology~\cite{doi2020cache}), similarly to the domain decomposition in parallel computing. In addition to divide the state vector, qubit reordering or remapping are used to decrease data exchanges between large numbers of qubit gates: these techniques require to insert swap gates in quantum circuits. 

\end{itemize}

The result of the Qiskit transpiler step is a new quantum circuit that has been optimized for performance. The new quantum circuit is then simulated by the simulator. Qiskit Aer provides a number of simulator approaches and supported hardware. In this work, we focus on GPU backends for the state vector simulator. In particular, Qiskit Aer provides two backends to enable execution on Nvidia GPUs (including multi-GPU):

\begin{itemize}
\item \textit{Nvidia Thrust.} The default GPU implementation in Qiskit uses Nvidia Thrust. Nvidia Thrust is a C++ template library that provides high-level parallel algorithms and data structures for programming GPUs~\cite{bell2012thrust}. It is part of the CUDA Toolkit, which is a set of tools and libraries provided by Nvidia for developing GPU-accelerated applications. Although Thrust allows developers to write code in C++ and leverage the power of GPUs without needing to directly write CUDA kernels or manage low-level GPU details. %

\item \textit{Nvidia cuQuantum.} Recently, Nvidia provides the cuQuantum SDK for optimized libraries and tools for accelerating quantum computing simulations~\cite{the_cuquantum_development_team_2023_7806810}. cuQuantum currently supports quantum circuit simulations based on state vector and tensor network methods and leverages Nvidia Tensor Core GPUs for speedup. The state vector cuQuantum library is a Qiskit Aer backend and can be easily executed. We note that cuQuantum is installed as binary library. 
\end{itemize}

\section{Methodology}\label{sec:methodology}
We develop the following suite of representative quantum applications and benchmarks to evaluate the GPU performance when running Qiskit Aer.

\noindent\textbf{\textit{Quantum Volume (QV) Circuit}}~\cite{cross2019validating} is an important benchmark circuit, composed of random instances of circuits, and used to measure the Quantum Volume metric (measure of the largest possible quantum circuit that a quantum computer can execute reliably). In particular, the QV benchmark circuit consists of layers of Haar random elements of the special unitary group SU(4) applied between corresponding pairs of qubits in a random bipartition. This bechmark takes as input the number of qubits and depth of the circuit.

\noindent\textbf{\textit{Quantum Fourier Transform (QFT)}}~\cite{shor1994algorithms} is the quantum analogue of the classical Fourier transform to convert a function in the time or spatial domain into its frequency domain representation. Probably the most important quantum algorithm at moments, the Shor's algorithm for the factorization into prime numbers, relies on QFT as a crucial step in factoring large numbers and solving the discrete logarithm problem efficiently on a quantum computer. The QFT building blocks Hadamard, controlled-rotation and swap gates. In our implementation, the QFT application takes the number of qubits as input.

\noindent\textbf{\textit{Quantum Random Circuit (QRC) Sampling}}~\cite{arute2019quantum} is a circuit proposed to demonstrate quantum computational supremacy: sampling from the output distribution of a large ($\approx > 50$ qubits) random quantum circuit is beyond the reach of classical computers, and for this reason, QRC serves as evidence for quantum computational supremacy. The QRC circuit consists of randomly chosen gates, including single-qubit gates and two-qubit gates, applied to a set of qubits. In this work, we use QASM circuits, generated by the the Google Cirq~\cite{isakov2021simulations} circuit used in~\cite{arute2019quantum}. The QRC circuit takes the number of qubits and circuit depth as the input.

\noindent\textbf{\textit{Grover's Circuit}}~\cite{grover1996fast} is another famous important quantum algorithm used for searching unsorted data. The basic building block of Grover's circuit is the so-called Amplitude Amplification (AA) primitive that allows to convert a phase difference in quantum states into an amplitude difference (therefore directly measurable). The AA primitive consists of Hadamard and phase gates. The second building block is the oracle circuit that encodes the information about the target solution that we want to find. In our benchmark, we use a simple oracle circuit that is implemented with a Toffoli gate ~\cite{adedoyin2018quantum}. The Grover's circuit takes as input the number of qubits.

\noindent\textbf{\textit{Greenberger-Horne-Zeilinger (GHZ) Circuit}}~\cite{greenberger1989going} is a circuit designed to create the so-called GHZ state that is a maximally entangled state: all qubits are in a superposition of being either all in the state $\ket{0}$ or all in the state $\ket{1}$. The GHZ state is used in quantum communication. The GHZ consists of a series of Hadamard gates and CNOT gates applied to a set of qubits. The input for the GHZ circuit is the number of qubits.

\noindent\textbf{\textit{1D Quantum Walk (QW)}}~\cite{montanaro2016quantum} is the quantum mechanics version of classical 1D random walks, e.g., a walker is placed in a one dimensional lattice and at each iteration a walker can shift either to the left or right lattice site with equal probability (in the quantum version this is implemented with a coin operator using an Hadamard gate). The beauty of quantum walks is they allow the walker to exist in a superposition of states and paths, making quantum interference effects occur. Instead of randomly choosing a path, a quantum walker follows a set of quantum operators that determine its motion. QWs are used for searching algorithms (similarly to Grover algorithms), quantum linear solvers, and quantum simulations. The basic implementation of QW circuits use right and shift operators (also known as increment and decrement primitives in quantum arithmetic) that are controlled by the coin operator, e.g., depending on the quantum state of the coin operator the walker shifts left or right. Differently, from the previous discussed quantum circuits, QW circuit depth depends on the number of iterations (or QW steps) and tends to be considerably large, even with a relatively small numbers of qubits and iterations.

\subsection{Experimental Setup}
\begin{table}[t]
    \caption{Hardware and software setup.}
    \centering
    \label{tab:exp_setup}
    \begin{tabular}{c|c}
        \toprule
        Setup & Details\\
        \midrule
         \textbf{CPU (\#)} & AMD EPYC (1)\\
         Cores   & 16\\
         Clock frequency   & 3.0 GHz\\
         Memory   & 128 GB DDR4\\
        \midrule
        \textbf{GPU (\#)} & Nvidia A100 (2)\\
        Memory per GPU & 40 GB HBM2 \\
        Theoretical peak memory BW per GPU & 1448 GiB/s \\
        Theoretical peak SP FLOPs per GPU & 10.5 TFLOP/s\\
        \midrule
        Compiler & GCC 8.5.0 \\
        MPI & OpenMPI 4.1.4 \\
        CUDA Toolkit & CUDA 11.5 \\
        Qiskit & 0.12 \\
        cuQuantum & 23.03.0 \\
        \bottomrule
    \end{tabular}
\end{table}
\begin{figure*}
    \centering
    \begin{subfigure}[t]{0.3\linewidth}
        \centering
        \includegraphics[width=\linewidth]{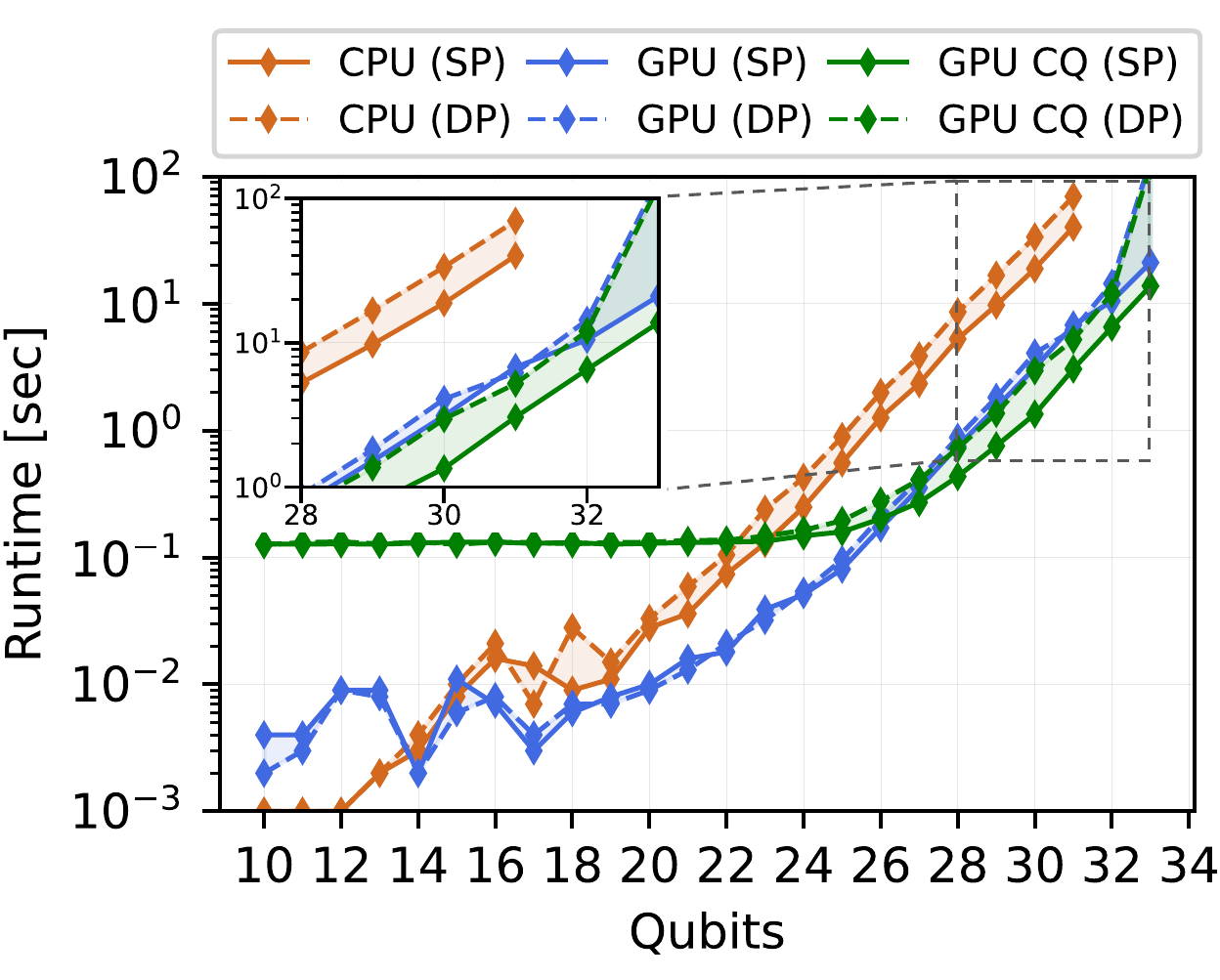}
        \caption{QV}
    \end{subfigure}
    \begin{subfigure}[t]{0.3\linewidth}
        \centering
        \includegraphics[width=\linewidth]{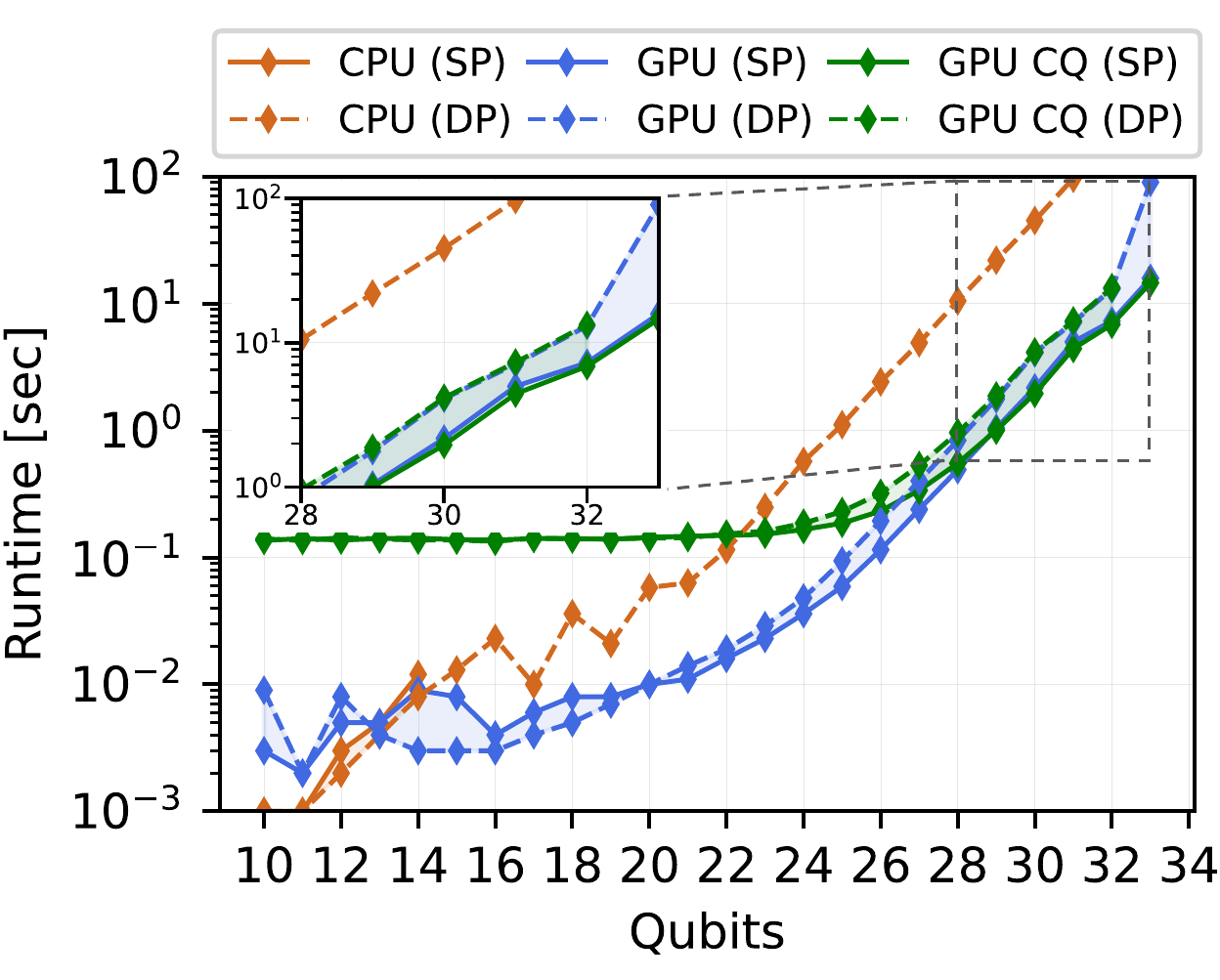}
        \caption{QFT}
    \end{subfigure}
    \begin{subfigure}[t]{0.3\linewidth}
        \centering
        \includegraphics[width=\linewidth]{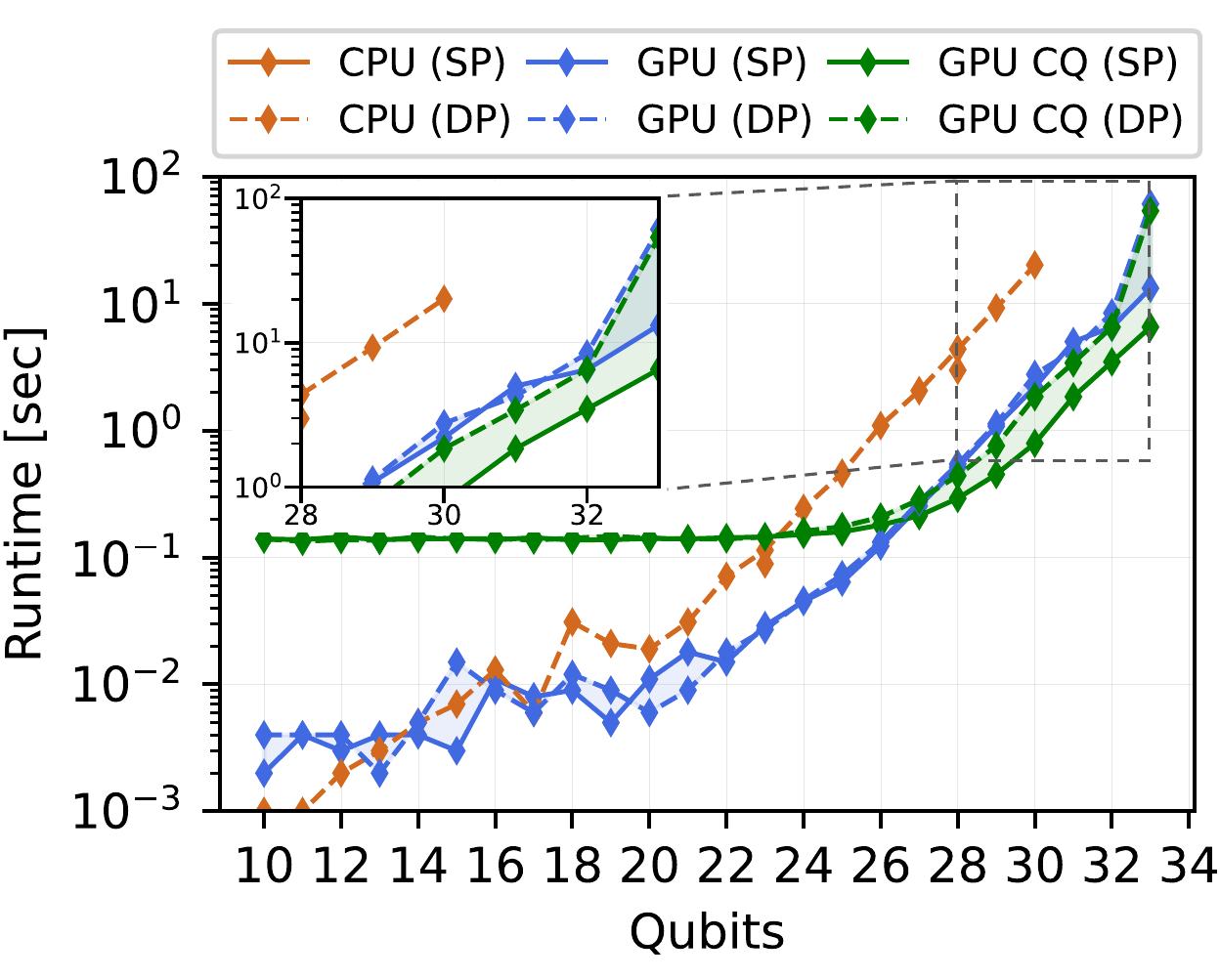}
        \caption{RQC}
    \end{subfigure}
    
    \begin{subfigure}[b]{0.3\linewidth}
        \centering
        \includegraphics[width=\textwidth]{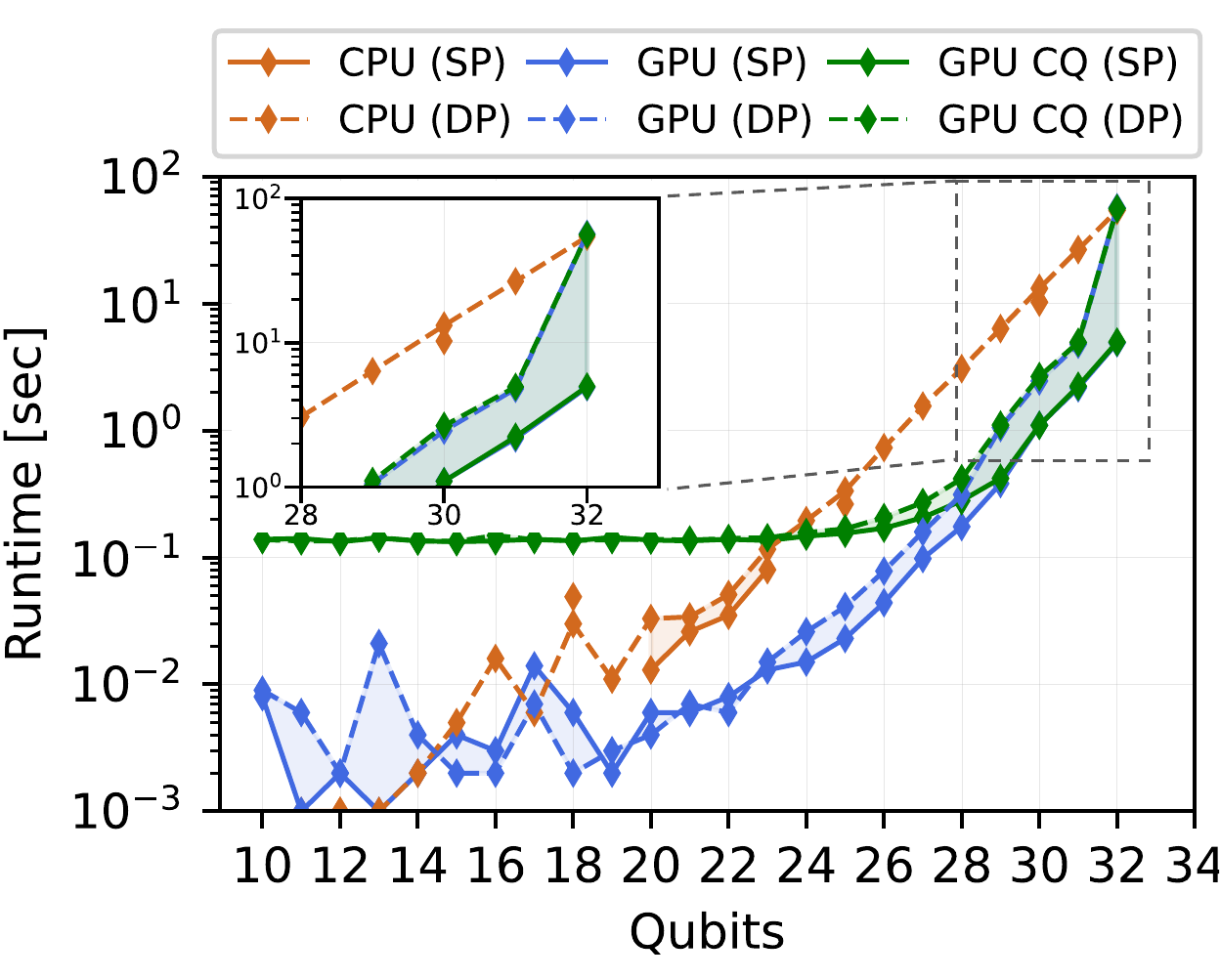}
        \caption{Grover}
    \end{subfigure}
    \begin{subfigure}[b]{0.3\linewidth}
        \centering
        \includegraphics[width=\textwidth]{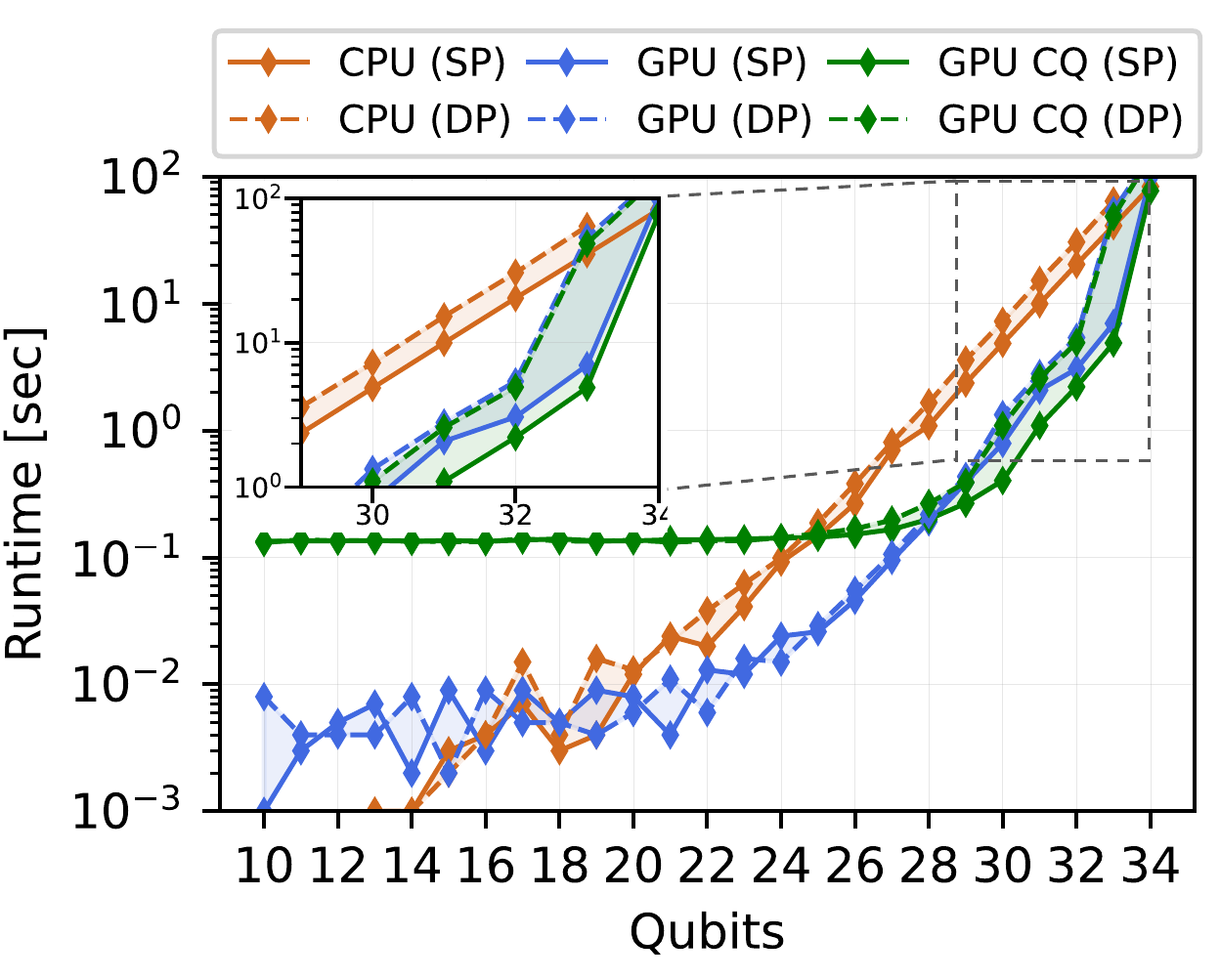}
        \caption{GHZ}
    \end{subfigure}
    \begin{subfigure}[b]{0.3\linewidth}
        \centering
        \includegraphics[width=\textwidth]{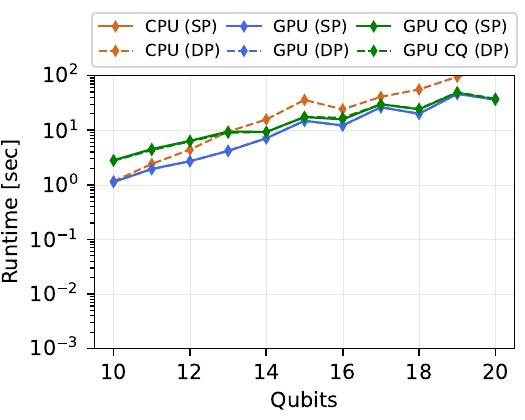}
        \caption{QW}
    \end{subfigure}
    \caption{Qiskit Aer execution time for the CPU, GPU with Thrust (denoted GPU), and GPU with cuQuantum  (denoted GPU CQ) backends when running the six applications using up to 34 qubits in single precision (SP) and double precision (DP). The usage of two GPUs allows to evaluate 31-34 qubits. The performance benefit of using GPUs is more than 10$\times$ for most of the application. For a large number of qubits, the cuQuantum backend for QV, RQC and GHZ is 1.5-3$\times$ faster than the Thrust GPU backend.\vspace{-1mm}}
    \label{fig:scaling}
\end{figure*}

\begin{table}[t]
    \caption{Benchmark setup for profiling.}
    \centering
    \label{tab:benchmarks_max}
    \begin{tabular}{c|c|cccc}
        \toprule
        \#GPUs &  & \#  &  & \#  & non-local\\ 
        (precision) & Benchmark & qubits & depth & gates & gates\\
        \midrule
         \multirow{6}{*}{1 (SP)} 
            & QV & 31 & 10 & 150 & 100\% \\
            & QFT & 31 & 62 & 511 & 94\%\\
            & RQC & 31 & 12 & 332 & 18\%\\
            & Grover & 30 & 11 & 67 & 3\%\\
            & GHZ & 31 & 31 & 31 & 98\%\\
            & QW & 16 (it=5) & 285712 & 351083 & 44\%\\
        \midrule
         \multirow{6}{*}{2 (DP)}
            & QV & 33 & 10 & 160 & 100\% \\
            & QFT & 33 & 66 & 577 & 94\% \\
            & RQC & 33 & 12 & 353 & 27\% \\
            & Grover & 32 & 12 & 71 & 3\% \\
            & GHZ & 34 & 34 & 34 & 97\% \\
        \bottomrule
    \end{tabular}
\end{table}

For the experiments, we use a multi-GPU system consisting of one AMD EPYC CPU and two Nvidia A100 GPUs. Details on the specifications of the system and the software environment are provided in Table~\ref{tab:exp_setup}. Since the focus of this work is the evaluation of GPU accelerators for state vector quantum computer simulators, we focus our profiling on the GPUs of the system, using Nvidia Nsight Systems \cite{nvidia_nsys} and Nsight Compute \cite{nvidia_ncu}. After performing scaling experiments, we utilize the Nvidia tools to obtain information regarding: 
\begin{itemize}
    \item distribution of the runtime between GPU kernels and other processes.
    \item GPU kernels occupying the largest portion of the time.
    \item roofline performance evaluation of each benchmark and the most used functions.
    \item memory footprints and data movement within each benchmark.
\end{itemize}

For each application, we chose a profiling setup that allows us to investigate \textit{(i)} single precision performance on a single GPU and \textit{(ii)} performance under full utilization of the system, i.e. using both GPUs and performing double precision computations. The characteristics of the chosen benchmark setups are shown in Table~\ref{tab:benchmarks_max}. For the single GPU evaluation, for each benchmark the largest number of qubits that was still executable on a single GPU was chosen. Blocking and, thus, multi-GPU execution was enabled in the system, i.e. with any qubit number larger than the one stated in the upper half of Table~\ref{tab:benchmarks_max}, two GPUs have been used automatically. Note that even the largest QW only uses single GPU. The circuit depths for each benchmark are either fixed (QV, RQC) or depend on the number of qubits (all others). In Table~\ref{tab:benchmarks_max}, we present also the total number of gates and the percentage of non-local gates (gates acting on one or more qubits, including controlled gates) \jennifer{after the circuit has been transpiled}.

For all applications, five experiments in sequence (the same circuit run five times in the same program) are carried out and we show the average execution time over the experiments. The variance of the execution time is negligible and therefore we omit the error bars when presenting the execution time.

\section{Evaluation}\label{sec:evaluation}
\begin{figure}[t]
    \centering
    \includegraphics[width=\linewidth]{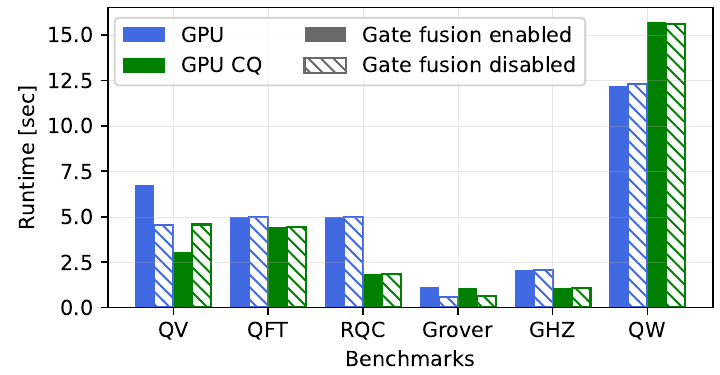}
    \caption{Impact of using gate fusion for the different quantum applications.\vspace{-1mm}}
    \label{fig:gatefusion}
\end{figure}

As a first step of our study, we study the execution time for CPU and GPU backends when increasing the number of qubits. Fig.~\ref{fig:scaling} shows the scalability of the state vector method on all six benchmarks from 10 to 34 qubits when utilizing either the CPU (orange color) or two GPUs. For the latter, both Qiskit's default Thrust backend (blue) and Nvidia's cuQuantum backend (green) are evaluated. Note that some of the benchmarks could only be run in double precision on the CPU because single precision runs lead to segmentation faults. It can be observed that, with a low number of qubits, the runtime of each benchmark is lower when utilizing the CPU. At a break point between 14 to 18 qubits, the GPU  with the Thrust backend becomes the fastest choice. The cuQuantum backend shows constant behaviour for up to $24-28$ qubits, and only afterwards becomes as fast as, or in some cases faster than the default Thrust backend of Qiskit. Specifically for the QV, RQC and GHZ benchmarks the cuQuantum backend is approximately $1.5-3\times$ faster than the Thrust backend for $28-32$ qubits. By investigating the trace, the nearly constant execution time for the cuQuantum backend for a small number of qubits, i.e., $10-22$, is due to the large overhead of memory allocation and pinning that dominate the execution time. 

Starting from 31 qubits in double precision, and 32 qubits in single precision, both GPUs on the system are utilized for both backends. Up to 34 qubits could be simulated in double precision on two GPUs for the GHZ benchmark. On CPU the same benchmark could be run with 34 qubits in single precision only. Here, the GPUs only reach a 1.1$\times$ speedup and even a 0.8$\times$ slowdown with cuQuantum and Thrust, respectively. For lower qubits a higher speedup is reached, e.g. for 30 qubits in single precision, the Thrust backend reaches a 6$\times$ speedup and the cuQuantum backend a 12$\times$ speedup. Similar behavior was found for the other benchmarks, where the GPUs generally reach a range between 32 and 33 qubits double precision, whereas the CPU could only simulate around 31 to 32 qubits for the same application, either not having sufficient memory available, or running for an exceedingly long time. For example, with this maximum number of qubits the GPU outperforms the CPU with up to a factor of around $13-14\times$ for both the QV and QFT benchmarks in double precision with cuQuantum and Thrust.

As the QW application is characterized by a large depth, overall a lower number of qubits was executable in reasonable time spans. Furthermore, up to the maximum number of 20 qubits still only a single GPU is utilized. With the default Thrust backend the single GPU execution time is more than twice as fast as the CPU runtime after 13 and more qubits. With 20 qubits, the GPU with Thrust backend outperforms the CPU by a factor of nearly four. The cuQuantum backend is not beneficial for this benchmark. While it is close to the default Thrust backend with a larger number of qubits, there is still a difference of around 3-5 seconds between these two.

As we discussed in the background section, the gate fusion performance optimization could lead to a performance boost, especially when GPUs are used. By default, in Qiskit Aer quantum state simulators, the gate fusion is applied to circuits with a number of qubits greater than 14. We study the performance impact of using gate fusion and present it in Fig.~\ref{fig:gatefusion}. We observe a big impact of gate fusion for QV with Thrust backend improving the performance in terms of execution time by a factor of 1.5$\times$ (conversely, the transpiler gate fusion slows down the execution time of the cuQuantum backend). For other applications, we do not observe major impact of gate fusion.
 \begin{figure}[t]
    \centering
    \begin{subfigure}[t]{0.15\textwidth}
        \centering
        \includegraphics[width=\linewidth]{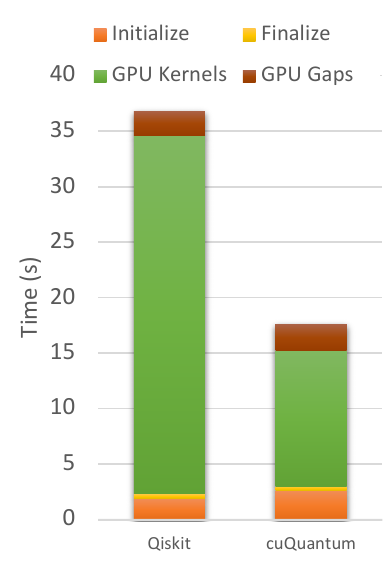}
        \caption{QV}
    \end{subfigure}
    \begin{subfigure}[t]{0.15\textwidth}
        \centering
        \includegraphics[width=\linewidth]{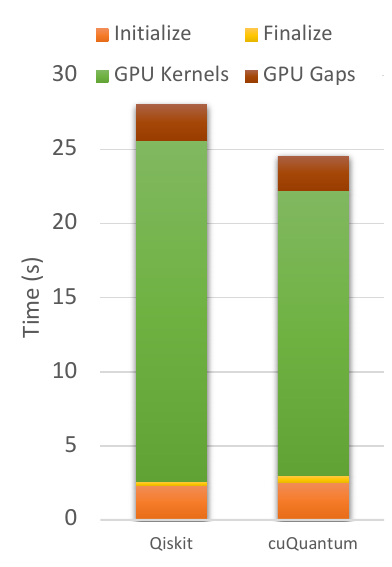 }
        \caption{QFT}
    \end{subfigure}
    \begin{subfigure}[t]{0.15\textwidth}
        \centering
        \includegraphics[width=\linewidth]{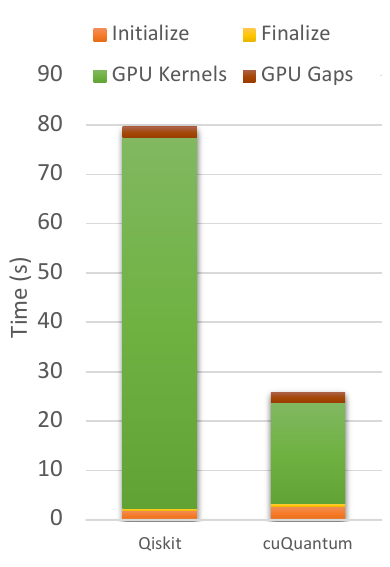 }
        \caption{RQC}
    \end{subfigure}
    \begin{subfigure}[t]{0.15\textwidth}
        \centering
        \includegraphics[width=\linewidth]{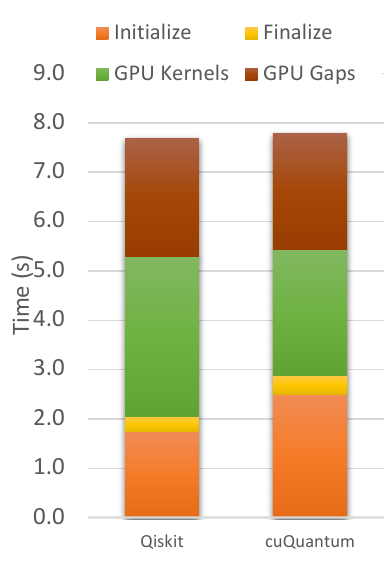}
        \caption{Grover}
    \end{subfigure}
    \begin{subfigure}[t]{0.15\textwidth}
        \centering
        \includegraphics[width=\linewidth]{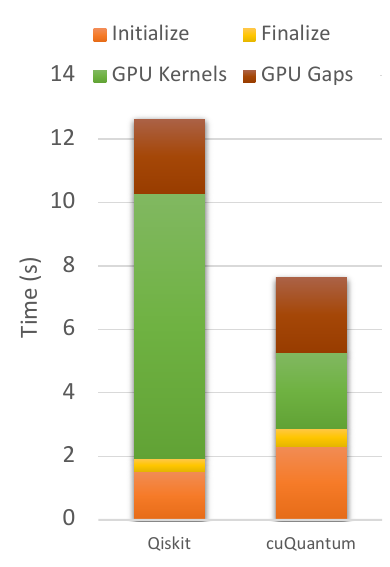}
        \caption{GHZ}
    \end{subfigure}
    \begin{subfigure}[t]{0.15\textwidth}
        \centering
        \includegraphics[width=\linewidth]{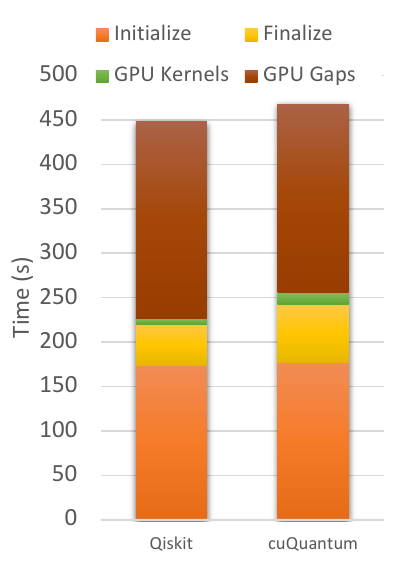}
        \caption{QW}
    \end{subfigure}    
    \caption{Breakdown of the different GPU activities on one GPU. The GPU running kernels (compute) are represented in green color. GPU Gaps refers to the periods with no GPU activities. Data transfer from/to host and device memories are negligible. The QW application is dominated by initialization part that requires to move the circuit configuration to the GPU (QW has a very deep quantum circuit).\vspace{-1mm}
    }
    \label{fig:gpu_breakdown}
\end{figure}
\begin{figure*}
    \centering
    \includegraphics[width=\linewidth]{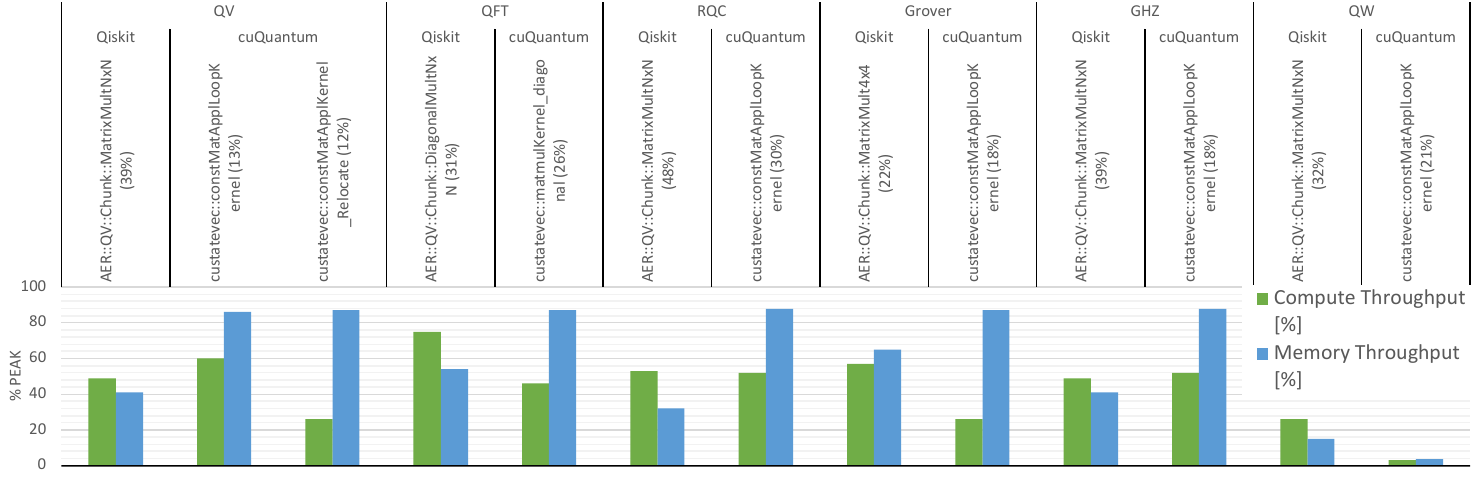}
    \caption{Most time consuming ($\geq$5\%) GPU compute kernels (shortened function names) when simulating largest number of qubits on a single GPU in single precision for each application. The compute and memory throughput, defined as the utilization of the theoretical their peak throughput, respectively, in the top kernels used in the Thrust and cuQuantum backends. The cuQuantum GPU backend stresses memory throughput in all top functions more than Qiskit Thrust backend, resulting in the blue bar higher than the green bar in all top functions. }
    \label{fig:throughput}
\end{figure*}

After studying the scalability of quantum applications, we investigate the GPU utilization and analyze the breakdown of the different activities. Fig.~\ref{fig:gpu_breakdown} presents compute activity (GPU kernels) in green, initialization in orange, finalization in yellow and GPU idle in brown. We can see that for the QV, QFT, and RQC benchmarks, the compute time dominates the execution time for both Thrust and cuQuantum backends, showing an efficient utilization of the GPU. The QW application is dominated by an initialization and GPU idle time, without an efficient usage of the GPU. Most interestingly, we see cuQuantum can accelerate GPU kernels (green) by $1.2-3.7\times$ compared to the Thrust backend in all applications except the QW benchmark.

To further understand how the quantum applications use the GPU compute units and memory system, we identify with Nvidia Nsight Systems the GPU kernels that take most time when running the two GPU backends and present them on the top of Fig.~\ref{fig:throughput}. We notice that matrix multiply are the dominant compute kernels when using the Thrust backend. Qiskit's Thrust backend provides custom implementation for multiply with matrix 2$\times$2, 4$\times$4 and 8$\times$8 and matrix multiply with generic $N\times N$ size. These large size matrix multiply are the result of the gate fusion optimization. %

We present the percentage of compute and memory throughput in these most used kernels in Fig.~\ref{fig:throughput}. The compute throughput is calculated as the percentage of ALU pipe in active cycles and memory throughput is calculated as the percentage of DRAM in active cycles. The cuQuantum backend shows considerably higher memory and compute throughput than the Thrust backend, except for the Grover and QW applications. We also notice that, compared to Thrust, the cuQuantum backend always tends to stress memory more than compute, i.e., all blue bars are higher than green bars.

\begin{figure}[t]
    \centering
    \includegraphics[width=\linewidth]{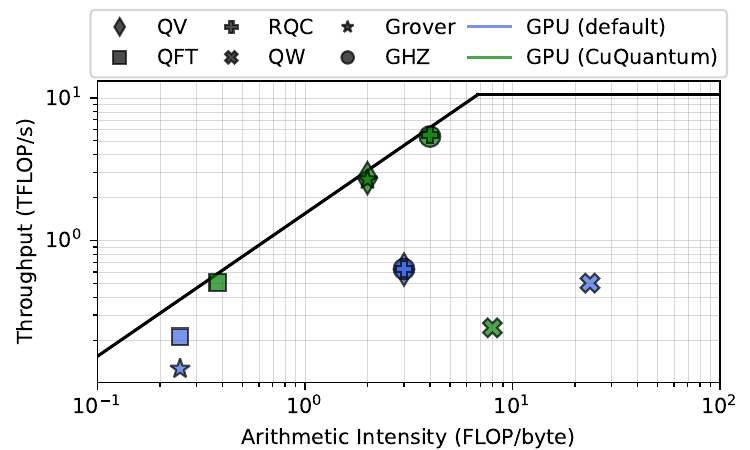}
    \caption{Single precision roofline of the A100 GPU and performance of the most used kernel of the six applications. The cuQuantum GPU backend reaches the roofline for all the quantum applications except for QW.}
    \label{fig:roofline}
\end{figure}

As last step in our performance characterization of the Qiskit GPU backends for state vector simulation, we present the roofline for the different kernels to determine whether the kernels are memory or compute bound in Fig.~\ref{fig:roofline}). Using the most utilized functions, we show the characteristics of each benchmark setup (Table~\ref{tab:benchmarks_max}, 1 GPU (SP)) in a roofline model of one A100 GPU, with the ridge point at an arithmetic intensity of 6.8, 1448 GiB/s peak memory bandwidth and 10.5 TFLOP/s single precision peak performance. The characteristics are obtained for the default backend (blue) and the cuQuantum backend (green). In Fig.~\ref{fig:roofline}, it can be seen that all applications fall in the memory bound region, except from the QW application that has low bandwidth and compute intensity. However, the performance reached with the latter still states a big gap towards the roofline. When comparing the default backend with the cuQuantum backend, cuQuantum utilizes the compute resources more efficiently and, thus, reaches the roofline for all the other benchmarks, explaining the performance improvement observed earlier in Fig.~\ref{fig:scaling}.

For state vector quantum circuit simulations, the memory usage for the state vector scales up exponentially with the number of qubits. Thus, single GPU simulation typically can only support simulations with qubits fewer than 32, as show in the previous experiments. We further scale up the scale of simulations by exploiting additional GPU memory in a multi-GPU setup. 

\begin{figure}[bt]
    \centering
    \includegraphics[width=\linewidth]{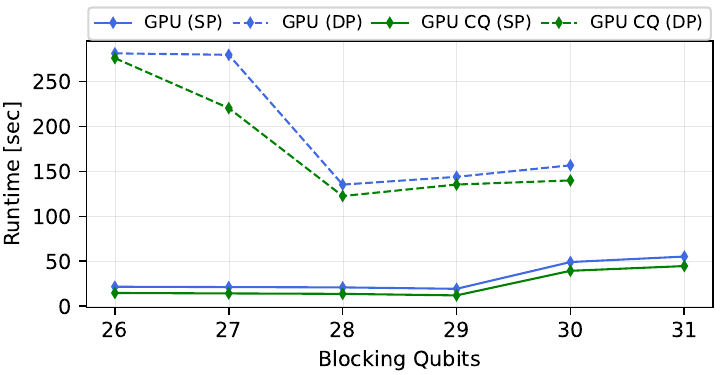}
    \caption{Runtime with a varying number of \texttt{blocking\_qubits} for a fixed number of 33 total qubits on 2 GPUs.}
    \label{fig:blocking}
\end{figure}
\begin{figure}[bt]
    \centering
    \includegraphics[width=\linewidth]{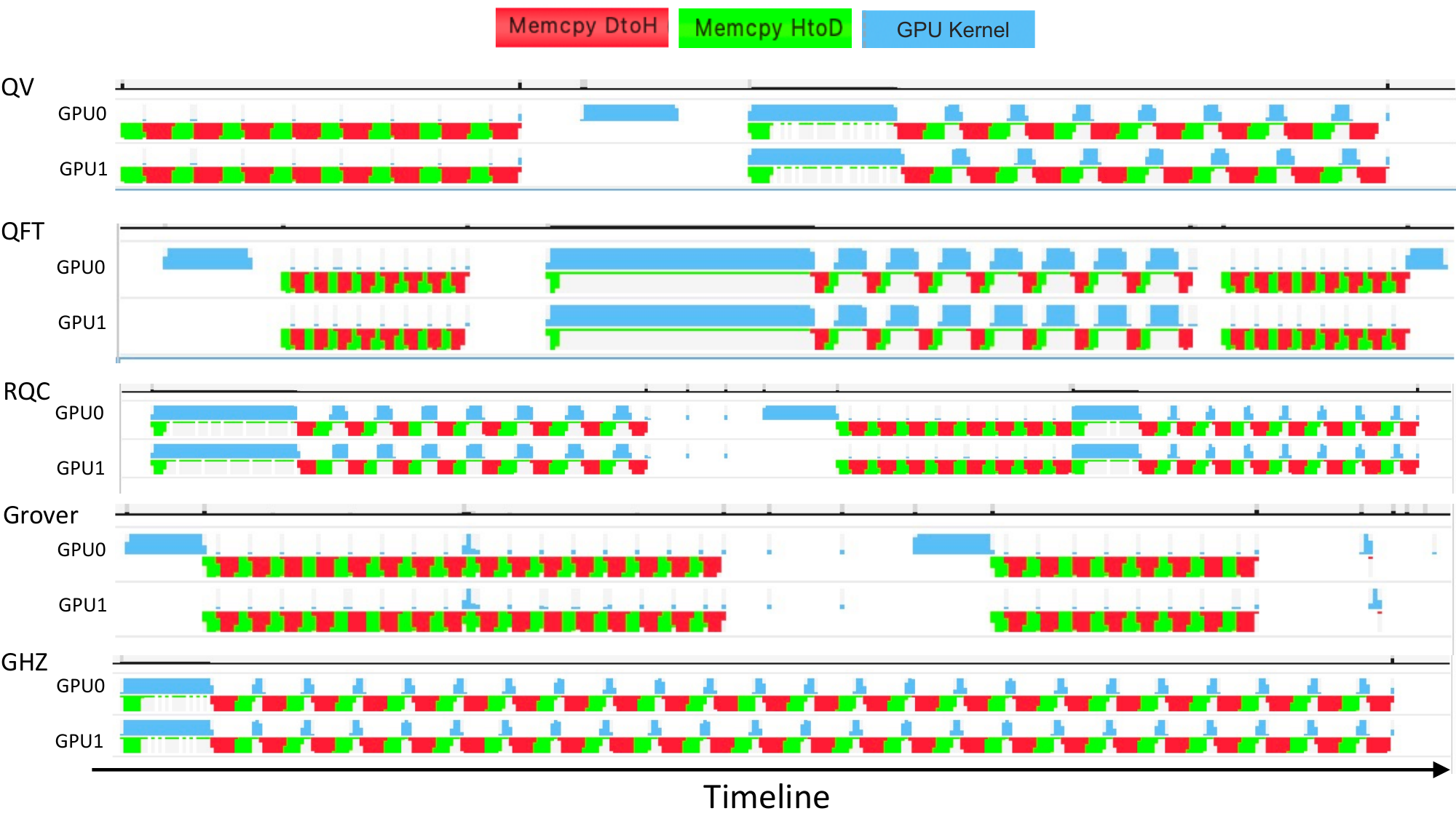}
    \caption{Trace of the data movement and computation for all the applications on two GPUs, except for QW. The typical pattern is a large data movement from host to device (H2D) and device-to-host (D2H) before and after GPU kernel execution with little to no overlapping.}
    \label{fig:timeline}
\end{figure}

When performing the simulation on multiple GPUs, the state vector is divided into different chunks and communication is needed between the memory of the two GPUs. The most important factor determining the data movement is the number of Qiskit \texttt{\small{blocking\_qubits}}. To showcase the impact of the \texttt{\small{blocking\_qubits}} on two GPUs, we perform experiments with the QV benchmark with 33 qubits and both GPU backends in single and double precision, varying the number of blocking qubits. The results shown in Fig.~\ref{fig:blocking} suggest that, while for single precision the difference between 26-29 is negligible, for double precision a blocking factor of 28 is most beneficial for both backends. The same was confirmed for other benchmarks and, thus, a factor of 28 was chosen for all other experiments.

When studying the performance on multi-GPU systems, it is critical to understand the memory footprint and data movement across the two GPU memories, as now the state vector is divided into different chunks residing on different memories. Investigating the memory footprint on two GPUs, the total memory usage is 72~GB for all the applications and the two GPU backends. 

Overall, data movement between host and GPUs becomes the top limiting factor of performance, taking more than $90\%$ GPU time in all two-GPU experiments. To understand the data movement, we inspect the tracing of the quantum applications. In Fig. \ref{fig:timeline}, we show the trace of five applications using two GPUs with the Thrust backend. The traces show one experiment running the quantum circuit for the applications with the largest possible number of qubits (see \ref{tab:benchmarks_max}), except for QW as it was only executed on a single GPU. Three colors represent three main phases: memory copies from host to device (green), compute (light blue) and memory copy from the device to the host (red color). By analyzing Fig.~\ref{fig:timeline}, we see that  data movement is similar for all the applications -- there is an extensive data movement from host to device (H2D) and device-to-host (D2H) before and after GPU kernel execution with little to no overlapping. Note that Fig.~\ref{fig:timeline} presents only a part of the whole timeline due to space limit.

\begin{figure}[bt]
    \centering
    \begin{subfigure}[t]{\linewidth}
        \centering
        \includegraphics[width=\linewidth]{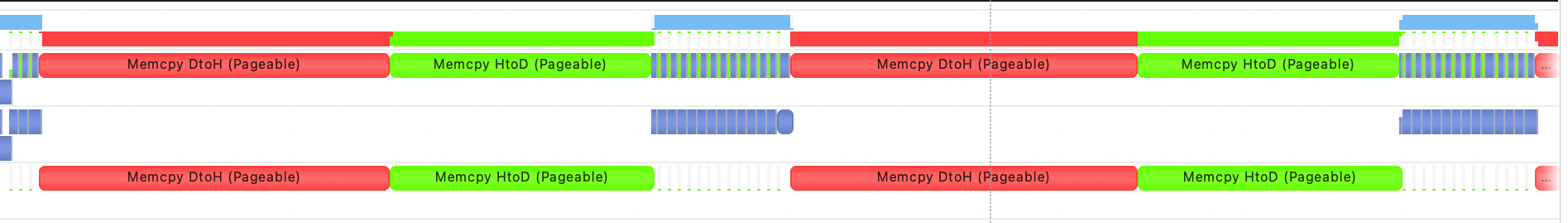}
        \caption{GPU Thrust backend.}
    \end{subfigure}
    \begin{subfigure}[t]{\linewidth}
        \centering
        \includegraphics[width=\linewidth]{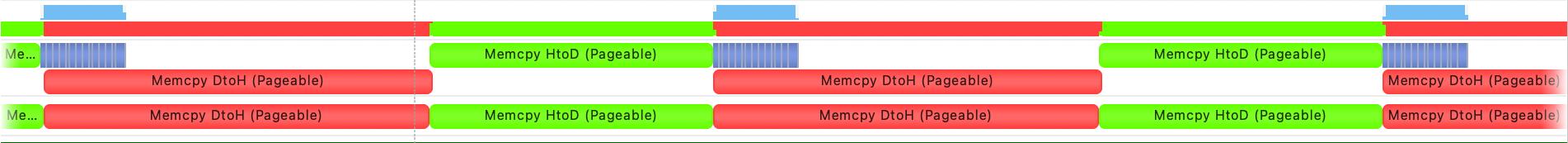}
        \caption{GPU cuQuantum backend.}
    \end{subfigure}    
    \caption{Comparison of memory transfer in the Thrust and cuQuantum backend multi-GPU strategy: cuQuantum overlaps kernel execution (blue) with \texttt{\small{cudaMemcpyAsync}} \jennifer{D2H} data transfer (red). However, since the kernels are much shorter than data transfers, no large performance improvement in end-to-end simulation time is observed.}
    \label{fig:datatransfer}
\end{figure}

In general, the traces obtained running the cuQuantum backend are similar to the ones with the Thrust backend. However, a main difference can been seen when comparing the traces. In Fig.~\ref{fig:datatransfer}, we show a zoom-in the traces for GHZ application comparing the the Thrust and cuQuantum backends. We can see that, for the data transfer, the cuQuantum backend uses \texttt{cudaMemcpyAsync} to overlap kernel execution (light blue) with device to host data transfer (red). However, because the kernels are much shorter than data transfers, no overall performance improvement is observed.
As we are using a multi-GPU node, we have the possibility of using MPI for dividing the workload between two GPUs. In an evaluation of distributed processing using MPI, we observe that the runtime increases significantly earlier when using MPI compared to the previous experiments that use an OpenMP based distribution across the two GPUs of our single node setup (Table~\ref{tab:mpi_nompi}). 
\begin{table}[tb]
    \caption{Runtime of the QV benchmark when utilizing two GPUs with the default Thrust backend, with and without MPI communication.}
    \centering
    \label{tab:mpi_nompi}
    \begin{tabular}{c|c|c|c}
        \toprule
        Qubits & Precision & Runtime [s] no MPI & Runtime [s] MPI \\
        \midrule
        \multirow{2}{*}{32} & SP & 10.456 & 15.194 \\
                            & DP & 14.349 & 108.060 \\
        \midrule
        \multirow{2}{*}{33} & SP & 21.041 & 149.562\\
                            & DP & 135.318 & 377.686\\
        \bottomrule
    \end{tabular}
\end{table}
\begin{figure}[bt]
    \centering
    \begin{subfigure}[t]{\linewidth}
        \centering
        \includegraphics[width=\linewidth]{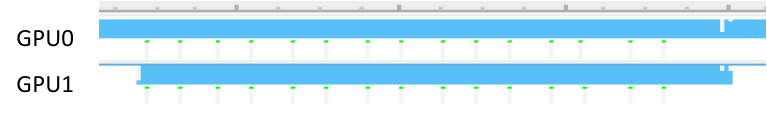}
        \caption{Without MPI}
    \end{subfigure}
    \begin{subfigure}[t]{\linewidth}
        \centering
        \includegraphics[width=\linewidth]{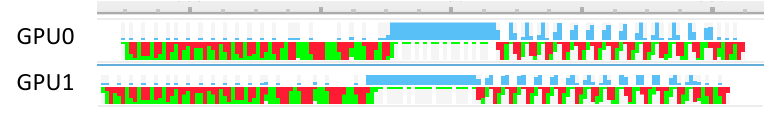}
        \caption{With MPI}
    \end{subfigure}
    \caption{Comparison of traces with and without MPI on the Quantum Volume benchmark with 33 qubits in single precision on 2 GPUs.}
    \label{fig:mpi}
\end{figure}

Fig.~\ref{fig:mpi} compares the traces with 33 qubits in single precision where the execution time is deviating significantly. It can be observed that the high \jennifer{H2D} and \jennifer{D2H} transfers are already present in the MPI version of this setup, while the non-MPI version reaches this stage with 33 qubits in double precision as shown earlier in Fig.~\ref{fig:timeline}.

\section{Related Work}
\label{sec:relatedwork}
State vector quantum computer simulators have a long history of development. The first parallel HPC quantum computer simulation approaches have been proposed by Da Raedt et al~\cite{daraedt2007parsim}, enabling the simulation of up to 36 qubits with the state vector representation, running benchmarks within 1,707 seconds when utilizing up to 4096 CPUs on the IBM BlueGene/L machine. Later on, Da Raedt et al~\cite{daraedt2019parsim} revised the massively parallel state vector simulator and provide additional performance measurements of up to 48 qubits on the Sunway TaihuLight and K supercomputers. 

Now, supercomputers have evolved to include different types of accelerators such as GPUs, whose utilization has been facilitated through high-level programming models, leading to the development of frameworks such as Qiskit that allow GPU-based quantum computer simulations. In this context, Doi et al.~\cite{doi2019} perform scaling experiments on a heterogeneous CPU-GPU setup, showing the applicability of such a setup for quantum computer simulations. Their simulator has been integrated into the Qiskit Aer framework. In a more recent work, Doi et al.~\cite{doi2020cache} implemented the cache-blocking technique that is used within Qiskit for multi-GPU acceleration. Using an IBM Power System AC922 with 6 GPUs, they simulate up to 35 qubits on a single node using the QV and QFT benchmarks, and perform further scaling experiments over multiple nodes. So far our presented work shows an in-depth analysis of the performance of this approach on a single node, leaving multi-node performance evaluation with a focus on MPI communication as a potential future work. 

Imamura et al.~\cite{imamura2022mpiqulacs} compare the aforementioned GPU performance to their CPU framework, mpiQulacs, focusing on distributed multi-node acceleration of the state vector simulator with MPI. They compare the runtime (double-precision) of their accelerator with Qiskit Aer on the Quantum Volume benchmark with the same circuit depth used in our setup running on a GPU cluster with 6 Nvidia V100 GPUs per node. In their setup, mpiQulacs outperforms Qiskit after scaling to four nodes with 32 qubits. However, for the same benchmark on our setup it was possible to scale well with up to 32 qubits on a single node with only two GPUs, requiring only 12 seconds, whereas the runtime on four CPU nodes with mpiQulacs lies above 20 seconds according to their results. This might be explained by the fact that there is no communication overhead in our setup since our evaluation is done on a single node only, which on the contrary speaks for a higher efficiency of the GPU setup over the CPU setup.

\section{Discussion \& Conclusion}
\label{sec:conclusion}
In this paper, we assessed the potential of using GPUs to simulate quantum computers using the state vector approach that simulates the evolution of the quantum state complex array after applying a number of gate transformations. We evaluated the performance of different quantum applications with the IBM Qiskit Aer simulator that provides two backends for GPUs: one based on Nvidia Thrust library and one based on the cuQuantum SDK. 

We are now in the position of answering the initial research question: \emph{are GPUs suitable and a key enabling technology for the acceleration of quantum computer simulations?} Brief answer: yes they are. Overall, we found that GPUs can provide a large improvement of performance with cuQuantum providing a major computational boost both in memory and compute throughput. In particular, for simulations with large number of qubits, GPUs can provide up to $14\times$ performance boost with Nvidia cuQuantum being $1.5-3\times$ faster than the original Thrust backend for several of the presented benchmarks and reaching the roofline in our experiments. While the usage of cuQuantum's compute capability are outstanding, we note that Nvidia tensor core units~\cite{markidis2018nvidia} are not used. Their usage could in principle provide an extra performance gain at the cost of reduced accuracy of the calculations (tensor cores work in mixed precision).

The major obstacle of state-vector simulations is that they quickly hit the memory and computational wall due to the exponential growth of the computational and memory requirements with the increase of the number of qubits we want to simulate. One possibility is to use multiple computational nodes or disaggregated systems and memory pooling~\cite{wahlgren2022evaluating} for additional memory. However, we have shown that at least on-node, the integration of MPI and GPU backends is not optimal. The usage of MPI is a necessary technology for high efficiency. Yet, the challenge of exponential memory will limit the state vector simulation below 50 qubits.

Two main approaches could address these limitations. The first strategy is to adopt hybrid approaches, combining for instance state vector and Feynman path simulator~\cite{markov2018quantum}, that trade memory usage for increased computation. For instance, this simulation approach is used in Google \texttt{qsimh} simulator~\cite{isakov2021simulations}. Because these hybrid approaches use the state vector simulator technology they are capable of exploiting GPU acceleration, as shown in this work. A second strategy is to exploit the sparsity of state vector and use compression. We note that the Qiskit Aer and many other state-of-the-art simulators use dense state vector. Important considerations for future work should be sparse representations, for instance like in bitwise-representation quantum computer simulators~\cite{da2020qsystem}, and usage of sparse linear algebra libraries for GPUs.

\section*{Acknowledgment}
Funded by the European Union. This work has received funding from the European High Performance Computing Joint Undertaking (JU) and Sweden, Finland, Germany, Greece, France, Slovenia, Spain, and the Czech Republic under grant agreement No 101093261.

\bibliographystyle{IEEEtran}
\bibliography{references}

\end{document}